\documentclass[12pt,epsfig]{article}
\usepackage{graphicx}
\usepackage[usenames]{color}
\usepackage{epsfig}
\usepackage{latexsym}
\definecolor{rotblass}{cmyk}{0,0.07,0.07,0.001}
 \newcommand\beq{\begin{equation}}
 \newcommand\eeq{\end{equation}}
 \newcommand\beqn{\begin{eqnarray}}
 \newcommand\eeqn{\end{eqnarray}}

\newcommand\bla{\color{black}}

\newcommand\ma{\color{magenta}}

\newcommand\red{\color{red}}
\definecolor{lg}{cmyk}{0,0,0,0.1}
\definecolor{lg2}{cmyk}{0,0,0,0.3}
\definecolor{la}{named}{Lavender}
\definecolor{lb}{named}{SkyBlue}
\definecolor{sg}{named}{SpringGreen}
\definecolor{do}{named}{BurntOrange}
\definecolor{or}{rgb}{1,0.8107,0.8077}

\title{
\begin{center}
{\Large\ma DESCRIPTION OF MULTIPARTICLE PRODUCTION}\\[0.2cm]
{\Large\ma BY GLUON DOMINANCE MODEL $^{1}$ }\\[0.2cm]
\end{center}
\vspace{1cm}
}
\author{{\Large\bla L.N.Abesalashvili, L.T.Akhobadze
  }\\[0.3cm]
{\large\sl\bla Institute High Energy Physics and Informatization, Tbilisi State 
 University, Georgia.} \\[0.3cm]
{\large\sl\bla E-mail: abesal@hepi.edu.ge }}
\date{ November 28, 2007}
\begin{document}
\maketitle
\topmargin 0cm
\oddsidemargin -0.4cm
\centerline{
                    {\Large\ma  ABSTRACT }}
\vspace{1.cm}		    
 The obtained $\pi^{-}$ and charged multiplicity distribution parameters of 
Gluon Dominance Model explain the experimental data in nucleus-nucleus, p nucleus,
pd, pp, p antip and $\pi^{-}$(p,n) interactions. We have undertaken an attempt to 
give description in different processes of multiparticle production by means of a 
unifed approach based on quark-gluon picture using the phenomenological 
hadronization. We have obtained agreement of GDM with experimental data in a very 
wide energy range.	   
   \par
\   \par
\   \par
\   \par
\   \par
\   \par
\   \par
\   \par
\   \par
\   \par
\   \par
\   \par
\   \par
{$^{1}$ Talk given at IV International Conference of Physics at the Future Colliders, IHEPI
TSU,October 22-26, Tbilisi, Georgia. }                      
\pagebreak
\par
Multiparticle production (MP) is one of the important branches in high energy 
physics [1].Modern accelerators had made it possible the intensive and detailed 
study of multiparticle processes.Multiplicity is the number of secondaries
$n$ in process MP:
\begin{center}
\begin{eqnarray}
\large {\red  
 A + B \rightarrow a_{1}+a_{2}+\cdots+a_{n}
\hspace{2cm}   }                     
\end{eqnarray}
\end{center}  
 Multiplicity distribution (MD) $P_{n}$ is the ratio cross-section 
\large {\red 
 $\sigma_{n}$ to $\sigma$=$\sum \limits_{n}$$\sigma_{n}$ 
$P_{n}$=$\sigma_{n}$/$\sigma$.
\hspace{1cm}   }
 To describe the MD we have used the probability of producing of $n$ charge 
particles in Gluon Dominance Model (GDM). GDM studies multiparticle production 
in lepton and hadron processes. It is based on the QCD and phenomenological 
scheme of hadronization.The model describes well multiplicity distributions 
and their moments.It has revealed an active role of gluons in multiparticle 
production, it also has confirmed the fragmentation mechanism of 
hadronization in e$^{+}$e$^{-}$ annihilation and its change to 
recombination mechanism in hadron and nucleus interactions.The GDM 
explains the shoulder structure of multiplicity distributions.The 
agreement with Au+Au peripheral collisions data for hadron-pion ratio
has been also obtained with this model. Development of GDM allows one
to study the multiplicity behavior of p+antip annihilation at tens of GeV.   
\par
 Heavy ion collisions (HIC) at high energies study strong evidences of
quark-gluon plasma (QGP) production [2].The behavior of bulk variables at
lower energies and also a detailed study of hadron interactions supply with 
understanding of the production mechanism of this new state. At present this 
analysis is realized at SPS (CERN) [3]. The basic problem of HIC is to describe 
the systems, consisting of partons or hadrons. Experiments at RHIC have 
confirmed this collective behavior [4]. In the case of the hadron interaction
the new formed medium, named quark-gluon plasma (QGP), won't have such a
plenty of constituents. We consider that the evaporation of single partons 
from separate hot pots (claster sources) in the system of coliding hadrons, leads 
to the secondary particles production. This conception was taken as the basis 
of the Gluon Dominance Model [5$\div$10]. It is supposed that after the 
inelastic collisions the part of the energy of the initial impact particles is 
transformed to the inside energy. Several quarks and gluons become free and form 
quark-gluon system (QGS). Partons which can produce hadrons are named the active 
ones. Two schemes were proposed [6,7]. In the first scheme the parton fission 
inside the QGS is taken into account (the scheme with a branch). If we are not 
interested in what is going inside QGS, we come to the scheme without a branch.
Reserve quarks remained inside of the leading particles. All of the newly born 
hadrons were formed by active gluons.
\par
 The Poisson distribution was chosen as the simplest multiplicity distribution 
for active gluons which appeared for the first time after the collision. The 
number of these gluons fulfils the role of the impact parameter for nucleus. On 
the second stage some of active gluons can leave QGS ("evaporate") and transform 
to real hadrons. For the hadronization a sub narrow binomial distribution (BD) 
was added as follows: 
\begin{center}
\begin{eqnarray}
\large {\red  
P_{n}=\sum\limits_{m=0}\limits^{M}C^{n-2}_{mN}(exp(-\overline{m})
\overline{m}^{m}/m!)(\overline{n}^h/N)^{n-2}(1-\overline{n}^{h}/N)^{mN-(n-2)},  
~(n>2),
\hspace{1cm}  }                                                     
\end{eqnarray}
\end{center}
\par
$(P_{2}$=exp(-$\overline{m}$)),
where $C^{n-2}_{mN}$ - binomial coefficient, $m$ and $\overline{m}$ are 
the number of secondary gluons and their mean multiplicites. In sum (2) 
we constrain the maximal possible number of the evaporated gluons equal 
to $M=6$. $\overline{n}^h$ and $N$ have the meaning of average multiplicity 
and a maximum possible number of secondary hadrons formed from the gluon at 
the stage of hadronization. 
\   \par
 The comparison (2) with experimental data [8, 11$\div$25] (see Fig.1$\div$9 and 
tables 6$\div$12), gives the following parameter values (see Tables 1$\div$5).
The expression (2) describes well the experimental data [8, 11$\div$25] from 
4 GeV/$c$ to 900 GeV (e.g. Fig. 1$\div$9 ). The mean gluon multiplicity 
$\overline{m}$ has a tendency to rise, but slower than the logaritmic one.
It is surprising that gluon parameters of hadronization ( $N$, $\overline{n}^h$ )
remain constant without considerable deviations in spite of the indirect finding: 
$N$$\sim$3$\div$4 and  $\overline{n}^h$$\sim$1. Therefore we can draw a conclusion 
about the universality of gluon hadronization in nucleus-nucleus collisions in the 
rather wide energy region.
\   \par
 As is shown by the analysis (2) gives better description of $\pi^{-}$(p,n), 
p(p,antip,d,nucleus)  and light nuclei collisions than for heavy nuclei.
\    \par
 In [26] MP is described by means of clan mechanism and emphasizes the gluon nature
of clan. GDM allows to give a concrete content for clan. The clan model uses the 
logaritmic distribution (LD) in a single clan.
\   \par
 At the SPS energy the shoulder structure appears in MD [13]. As it was mentioned in 
the branch scheme, the gluon fission is strengthened at higher energies. The 
independent evaporation of gluon sources of hadrons may be realized as single gluons
as groups from two and more fisson gluons. Following [26] such groups is named clans.
\   \par 
 The specific feature of GDM is the dominance of active gluons in MP. We expect the 
emergence of many of them in nucleus collisions at RHIC and the formation of a new kind 
of matter (QGP) at high energy. The QGS can be a candidate for this. According to GDM, 
the active gluons are a basic source of secondary hadrons.
\   \par
 In conclusion one can show GDM may explain experimental data [8, 11$\div$25] 
(see Fig. 1$\div$9 and tables 6$\div$12) end gives the following parameter values 
(see Tables 1$\div$5).
\   \par
 GDM describe well MD pp interactions at the region of (50$\div$800 GeV/c, 62 GeV)
(see Table 3 and fig.5), ($\pi^{-}$,p) interactions at the region of (40$\div$360 GeV/c
(see Table 2 and fig 4). The maximum possible number of secondary hadrons formed from  
the active gluons N, their mean multiplicity $\bar{n}^{h}$ increase slowly. A growth of
$\bar{n}^{h}$ in pp interactions indicates a possible change mechanism of hadronization
of gluons in comparison with (p antip) annihilation (see Table 4 and figure 6).
\   \par 
 The parameter of hadronization ($\bar{n}_{ch})^{h}$ has a tendency to increase weakly. We 
consider that parameter ($\bar{n}_{ch})^{h}$ goes to the limiting value (like saturation). 
For hadron and nucleus processes a lot of quark pairs from gluons appear almost 
simultaneously and recombine to various hadrons [27]. The value $\bar{n}^{h}$ becoms 
bigger $\sim$(2$\div$3), that indicates to the transition from the fragmentation 
mechanism to the recombination one.
\   \par
 In our research we see that: 
\   \par 
\large {\red 
 1. At the same energy the mean multiplicity of the active gluons $\overline{m}$,
the maximal possible number of secondary hadrons formed from one active gluons
at the second stage N and their mean multiplicity $\bar{n}^{h}$ is higher in the 
nucleus-nucleus collisions and annihilation processes, than in the hadron-hadron 
interactions.
\   \par
 2. With the growth of the energy of colliding particles the mean multiplicity of 
the active gluons $\overline{m}$ increase slowly in all interactions.
\   \par
 We have obtained agreement of Gluon Dominance Model (GDM) with experimental data 
in (p antip) annihilation, pp, ($\pi^{-}$,(p,n)), pd and nucleus-nucleus collisions in a 
very wide energy domain.
\   \par
 The specific feature of GDM is the dominance of active gluons in MP. We expect the
emergence of many of them in nucleus collisions and the formation of a new kind of 
matter (QGP) at high energy.
\hspace{1cm}   }
\   \par
 The authors appreciate for the support of physicists from HEPI TSU who encouraged 
our investigation.
\pagebreak
\par

\par
\newpage
\begin{center}
\bf{FIGURE CAPTIONS}
\end{center} \  \par
{\bf{Fig.1.}}
The multiplicity  distributions of  $\pi^{-}$ 
mesons in ((He,d,C),Ta) collisions at 4.2 GeV/$c$/nucleon.
The curves are the result of the approximation of experimental data
by sum (2) of Gluon Dominance Model.
\   \par
\par
{\bf{Fig.2.}}
The multiplicity  distributions of  $\pi^{-}$ 
mesons in ((He,C),C) collisions at 4.2 GeV/$c$/nucleon.
The curves are the result of the approximation of experimental data
by sum (2) of Gluon Dominance Model.
\   \par
\par
{\bf{Fig.3.}}
The multiplicity distributions of  $\pi^{-}$ mesons in ((He,C),Prop)
 collisions at 4.2 GeV/$c$/nucleon.
The curves are the result of the approximation of experimental data
by sum (2) of Gluon Dominance Model.
\   \par
\par
{\bf{Fig.4.}}
The multiplicity distributions of charged particles in 
($\pi^{-}$,p)$\rightarrow$(ch,X) at (40, 50, 205 and 360) GeV/$c$.
The curves are the result of the approximation of experimental data
by sum (2) of Gluon Dominance Model.
\   \par
\par
{\bf{Fig.5.}}
The multiplicity  distributions of charged particles in 
(p,p)$\rightarrow$(ch,X) at (50,300 and 800) GeV/$c$.
The curves are the result of the approximation of experimental data
by sum (2) of Gluon Dominance Model.
\   \par
\par
{\bf{Fig.6.}}
The multiplicity distributions of charged particles in 
(p,antip)$\rightarrow$(ch,X) at (14.75 and 22.4) GeV/$c$.
 The curves are the result of the approximation of experimental data
by sum (2) of Gluon Dominance Model.
\   \par
\par
{\bf{Fig.7.}}
The multiplicity distributions of charged particles ($\pi^{-}$,n)$\rightarrow$(ch,X) 
 at the momentum of 40 GeV/$c$/nucleon and
 (p,(p,n,d))$\rightarrow$ (ch,X) at the momentum of 300 GeV/$c$/nucleon.
The curves are the result of the approximation of experimental data
by sum (2) of Gluon Dominance Model.
\   \par
\par
{\bf{Fig.8.}}
The multiplicity  distributions of $\pi^{-}$ mesons 
 in (p,(Ar,Xe))$\rightarrow$($\pi^{-}$,X) 
 collisions at 200 GeV/$c$/nucleon.
The curves are the result of the approximation of experimental data
by sum (2) of Gluon Dominance Model.
\   \par
\par
{\bf{Fig.9.}}
The multiplicity  distributions of charged particles  
 in (p,(Ar,Xe))$\rightarrow$(ch,X) 
 collisions at 200 GeV/$c$/nucleon.
 The curves are the result of the approximation of experimental data
by sum (2) of Gluon Dominance Model.
\newpage
\begin{center}
Table 1. Parameters of gluon dominance model (GDM) \\
(He,d,C)Ta$\rightarrow$($\pi^{-}$,X),(C,Ta)$\rightarrow$(ch,X), 
(He,C)C$\rightarrow$($\pi^{-}$ ,X) and \\
(He,C)Prop$\rightarrow$ ($\pi^{-}$,X) at the momentum of 4.2
GeV/$c$/nucleon. \\
$A_{P}$ Projectile and $A_{T}$ target 
\end{center}
\begin{tabular}{|l|c|c|c|c|c|}    \hline
$A_{P},A_{T}$ & $\bar{m}$ & N & $\bar{n}^{h}$ & $\chi^{2}$/ndf &
$\chi^{2}$/$N_{exp}$  \\
\hline
 $(He,Ta)$& 2.34$\pm$0.12& 10.21$\pm$2.54& 3.99$\pm$0.49& 17/5& 17/8   \\
\hline
$(d,Ta )$& 2.82$\pm$0.22 &12.50$\pm$4.92& 4.42$\pm$0.69& 10/3&10/6   \\
\hline
 (C,Ta)$\rightarrow$($\pi^{-}$,X) & 2.81$\pm$0.08 & 5.03$\pm$0.27 & 1.97$\pm$0.10 & 8/6 & 11/6 \\
\hline
 (C,Ta)$\rightarrow$ (ch,X) & 3.84$\pm$0.13 & 5.92$\pm$0.38 & 2.23$\pm$0.13 & 14/12 & 14/15  \\ 
\hline
 $(He,C )$  & 2.89$\pm$0.21 & 10.23$\pm$3.04 & 5.40$\pm$0.49 & 11/3 & 11/6 \\
\hline
 $(C, C )$  &  2.34$\pm$0.13 & 12.40$\pm$5.90 & 5.14$\pm$1.49 & 15/5 &  15/8  \\
\hline
 $(He,Prop)$ & 1.93.67$\pm$0.31 & 10.07$\pm$4.28 & 5.74$\pm$1.96 & 5/3 &  5/6  \\
\hline
 $(C, Prop)$ & 2.86$\pm$0.12 & 12.49$\pm$5.26 & 5.13$\pm$2.01& 14/5 & 14/8 \\
\hline
\end{tabular}
\   \par
\   \par
\begin{center}
 Table 2. Parameters of gluon dominance model (GDM) 
 ($\pi^{-}$,p)$\rightarrow$ (ch,X) \\
 at the momentum of (40, 50, 205 and 360) GeV/$c$.  \\
 $A_{P}$ Projectile $\pi^{-}$ and $A_{T}$ target p
\end{center} 
\begin{tabular}{|l|c|c|c|c|c|}    \hline
$A_{P},A_{T}$ & $\bar{m}$ & N & $\bar{n}^{h}$ & $\chi^{2}$/ndf &
$\chi^{2}$/$N_{exp}$  \\
\hline 
   40 GeV/$c$ & 1.41$\pm$0.10 & 2.00$\pm$0.01 &1.36$\pm$0.06 & 6/6 & 6/9  \\
\hline
   50 GeV/$c$ & 2.61$\pm$0.68 & 1.00$\pm$0.01 &0.74$\pm$0.20 & 3/5 & 3/8   \\
\hline
  205 GeV/$c$ & 4.29$\pm$0.12 & 2.84$\pm$0.43 &0.81$\pm$0.15 & 10/7 & 10/10  \\
\hline
  360 GeV/$c$ & 4.34$\pm$0.17 & 4.47$\pm$1.22 &0.90$\pm$0.02 & 4.4/8 & 4.4/11 \\
\hline
\end{tabular}
\   \par
\   \par
\begin{center} 
Table 3. Parameters of gluon dominance model (GDM) \\
(p,p)$\rightarrow$(ch,X) at (50,200,205,300,400 and 800) GeV/c and at 62 GeV.\\
$A_{P}$ Projectile p and $A_{T}$ target p
\end{center}
\begin{tabular}{|l|c|c|c|c|c|}    \hline
$A_{P},A_{T}$ & $\bar{m}$ & N & $\bar{n}^{h}$ & $\chi^{2}$/ndf &
$\chi^{2}$/$N_{exp}$  \\
\hline      
  50 GeV/$c$ & 2.35$\pm$0.60 & 2.00$\pm$0.17 & 1.46$\pm$0.09 & 6/4 & 6/7 \\
\hline       
 200 GeV/c & 3.13$\pm$0.26 & 1.91$\pm$0.25 & 0.97$\pm$0.06 & 14/7 & 14/10 \\
\hline       
 205 GeV/c & 2.91$\pm$0.21 & 2.01$\pm$0.16 & 1.01$\pm$0.05 & 16/8 & 16/11 \\
\hline       
 300 GeV/c & 3.35$\pm$0.29 & 5.27$\pm$3.03 & 1.21$\pm$0.09 & 13/9 & 13/12 \\
\hline       
 400 GeV/c & 2.24$\pm$0.91 & 2.28$\pm$0.11 & 1.31$\pm$0.06 & 19/12 & 19/15 \\
\hline       
 800 GeV/c & 2.66$\pm$0.16 & 2.25$\pm$0.07 & 1.29$\pm$0.04 & 16/13 & 16/16  \\
\hline       
  62 GeV & 2.33$\pm$0.11 & 3.23$\pm$0.17 & 1.95$\pm$0.08 & 24/15& 24/18 \\
\hline       
\end{tabular}
\newpage
\begin{center}
Table 4. Parameters of gluon dominance model (GDM) \\
(p,antip)$\rightarrow$ (ch,X) at the momentum of (14.75, 22.4) GeV/c and at (200, 900) GeV. \\
$A_{P}$ Projectile p and $A_{T}$ target antip
\end{center}
\begin{tabular}{|l|c|c|c|c|c|}    \hline
&  &  &  &  &  \\
$A_{P},A_{T}$ & $\bar{m}$ & N & $\bar{n}^{h}$ & $\chi^{2}$/ndf &
$\chi^{2}$/$N_{exp}$  \\
&  &  &  &  &  \\
\hline            
  14.7 GeV/$c$ & 2.20$\pm$0.18 & 2.05$\pm$0.10 & 1.21$\pm$0.04 & 1/9 & 1/12 \\
\hline       
  22.4 GeV/$c$ & 2.26$\pm$0.18 & 2.01$\pm$0.05 & 1.33$\pm$0.05 & 8/8 & 8/11 \\
\hline       
 200 GeV & 3.59$\pm$0.17 & 3.00$\pm$0.20 & 1.67$\pm$0.05 & 45/16 & 45/19 \\
\hline       
 900 GeV & 5.68$\pm$0.01 & 4.11$\pm$0.70 & 1.27$\pm$0.01 & 13/8 & 13/11 \\
\hline       
\end{tabular}
\   \par
\   \par
\begin{center} 
Table 5. Parameters of gluon dominance model (GDM) \\
($\pi^{-}$,n)$\rightarrow$ (ch,X) at the momentum of 40 GeV/$c$, \\
(p,(Ar,Xe))$\rightarrow$(ch,X), (p,(Ar,Xe))$\rightarrow$($\pi^{-}$,X) at the momentum of 200 GeV/$c$ \\ 
 and (p,(n,d))$\rightarrow$ (ch,X) at the momentum of 300 GeV/$c$/nucleon. \\
 $A_{P}$ Projectile and $A_{T}$ target 
\end{center}
\begin{tabular}{|l|c|c|c|c|c|}    \hline
&  &  &  &  &  \\
$A_{P},A_{T}$ & $\bar{m}$ & N & $\bar{n}^{h}$ & $\chi^{2}$/ndf &
$\chi^{2}$/$N_{exp}$  \\
&  &  &  &  &  \\
\hline                       
 ($\pi^{-}$,n)$\rightarrow$(ch,X)40 & 1.45$\pm$0.12 & 2.71$\pm$0.13 &1.67$\pm$0.13& 6/5 & 6/8 \\
\hline                       
 (p,n)$\rightarrow$(ch,X) 300 & 2.44$\pm$0.13 & 2.41$\pm$0.11 & 1.40$\pm$0.01 & 13/9 & 13/12 \\
\hline                      
 (p,d)$\rightarrow$(ch,X) 300 & 2.65$\pm$0.12 & 3.33$\pm$0.15 & 2.14$\pm$0.01 & 10/9 & 10/12 \\
\hline                       
 (p,Ar)$\rightarrow$(ch,X) 200 & 3.40$\pm$0.10 & 5.00$\pm$0.43 & 3.27$\pm$0.01 & 43/17 & 43/20 \\
\hline
 (p,Xe)$\rightarrow$(ch,X) 200 & 5.64$\pm$0.33 & 4.83$\pm$0.02 & 3.19$\pm$0.03 & 58/11 & 58/14 \\
\hline 
 (p,Ar)$\rightarrow$($\pi^{-}$,X)200 & 2.25$\pm$0.14 & 2.51$\pm$0.14 & 1.58$\pm$0.1 & 31/19 & 31/22 \\                                                                         & \\
\hline                                                      
 (p,Xe)$\rightarrow$($\pi^{-}$,X)200&2.32$\pm$0.11&3.00$\pm$0.02& 1.98$\pm$0.02 & 31/19 &31/22 \\                                                                         & \\
\hline                       
\end{tabular}
\newpage
\begin{center} 
Table 6. Experimental results the multiplicity  distributions of  $\pi^{-}$ mesons \\
and of charged particles in ((He,d,C)Ta) collisions at 4.2 GeV/$c$/nucleon [20] \\
\end{center}
\begin{tabular}{|l|c|c|c|c|c|c|}    \hline
&  &  &  &  &  &  \\
Mult&    (d,Ta)   &   (He,Ta)    &   (C,Ta)    &(C,Ta)$\rightarrow$(ch,X)&Mult&  (C,Ta) \\
&  &  &  &  &  &  \\
\hline
 n & Pn$\pm$dPn  & Pn $\pm$dPn  & Pn $\pm$dPn &  Pn$\pm$dPn & n & Pn $\pm$dPn \\
\hline
 0 &.384$\pm$.02 &.274$\pm$.032 &.174$\pm$.026&.053$\pm$.01 &24 &.017$\pm$.005 \\                                                                           
 1 &.376$\pm$.02 &.341$\pm$.022 &.194$\pm$.015&.094$\pm$.01 &25 &.012$\pm$.005 \\
 2 &.185$\pm$.02 &.208$\pm$.018 &.122$\pm$.012&.087$\pm$.01 &26 &.010$\pm$.004 \\
 3 &.048$\pm$.01 &.124$\pm$.014 &.100$\pm$.01 &.068$\pm$.009&27 &.009$\pm$.004 \\
 4 &.004$\pm$.002&.043$\pm$.012 &.083$\pm$.01 &.054$\pm$.008&28 &.013$\pm$.004 \\
 5 &.002$\pm$.001&.006$\pm$.0025&.083$\pm$.01 &.046$\pm$.008&29 &.009$\pm$.004 \\          
 6 &.001$\pm$.001&.003$\pm$.0015&.071$\pm$.01 &.034$\pm$.007&30 &.011$\pm$.004 \\
 7 &        0.   &.001$\pm$.001 &.066$\pm$.008&.026$\pm$.007&31 &.013$\pm$.004 \\          
 8 &        0.   &              &.049$\pm$.008&.029$\pm$.007&32 &.007$\pm$.004 \\               
 9 &        0.   &              &.015$\pm$.007&.029$\pm$.007&33 &.007$\pm$.004 \\
 10&             &              &.016$\pm$.007&.028$\pm$.007&34 &.011$\pm$.004 \\
 11&             &              &.011$\pm$.005&.019$\pm$.007&35 &.008$\pm$.004 \\                        
 12&             &              &.009$\pm$.005&.025$\pm$.006&36 &.010$\pm$.004 \\
 13&             &              &.004$\pm$.003&.023$\pm$.006&37 &.008$\pm$.004 \\
 14&             &              &.009$\pm$.008&.023$\pm$.006&38 &.006$\pm$.003 \\
 15&             &              &.001$\pm$.001&.019$\pm$.005&39 &.009$\pm$.004 \\
 16&             &              &             &.017$\pm$.005&40 &.006$\pm$.004 \\
 17&             &              &             &.023$\pm$.005&41 &.008$\pm$.004 \\
 18&             &              &             &.015$\pm$.005&42 &.003$\pm$.002 \\
 19&             &              &             &.021$\pm$.005&43 &.008$\pm$.004 \\
 20&             &              &             &.012$\pm$.005&44 &.005$\pm$.003 \\ 
 21&             &              &             &.009$\pm$.005&45 &.004$\pm$.002 \\
 22&             &              &             &.014$\pm$.005&46 &.004$\pm$.002 \\
 23&             &              &             &.010$\pm$.005&47 &.003$\pm$.003 \\
\hline                                                        
\end{tabular}
\   \par
\newpage
\begin{center} 
Table 7. Experimental results the multiplicity  distributions of  $\pi^{-}$ mesons \\
  in (C,C), ((C,He)Prop) and (He,C) collisions at 4.2 GeV/$c$/nucleon [21] \\
\end{center}
\begin{tabular}{|l|c|c|c|c|}    \hline
&  &  &  &  \\
Mult&    (C,C)      &     (C,Prop)   &   (He,Prop)  &   (He,C)    \\ 
&  &  &  &  \\
\hline
 n & Pn $\pm$ dPn   &  Pn $\pm$ dPn  & Pn $\pm$ dPn & Pn $\pm$ dPn  \\
\hline
 0  & 0.177$\pm$0.08  &  0.417$\pm$0.04   & 0.482$\pm$0.055 & 0.349$\pm$0.018  \\
 1  & 0.375$\pm$0.04  &  0.327$\pm$0.025  & 0.369$\pm$0.045 & 0.419$\pm$0.028  \\
 2  & 0.256$\pm$0.03  &  0.153$\pm$0.015  & 0.108$\pm$0.035 & 0.166$\pm$0.017  \\
 3  & 0.097$\pm$0.02  &  0.053$\pm$0.008  & 0.030$\pm$0.02  & 0.048$\pm$0.009  \\
 4  & 0.063$\pm$0.02  &  0.034$\pm$0.005  & 0.01 $\pm$0.007 & 0.015$\pm$0.005  \\
 5  & 0.024$\pm$0.006 &  0.013$\pm$0.003  & 0.002$\pm$0.0017& 0.003$\pm$0.0015 \\
 6  & 0.005$\pm$0.0025&  0.0028$\pm$0.0015&     0.0         &                  \\
 7  & 0.003$\pm$0.0015&  0.0014$\pm$0.0007&     0.0         &                  \\
\hline
\end{tabular} 
\   \par
\newpage
\begin{center} 
Table 8. Experimental results the multiplicity  distributions of charged particles \\  
  in (($\pi^{-}$,n)) collisions at 40 GeV/$c$ [15,17-18], ($\pi^{-}$,p) collisions \\
  at 40 GeV/$c$ [15,17-18] and at (50,205 and 360)GeV/$c$ [22] \\
\end{center}
\begin{tabular}{|l|c|c|c|c|c|c|}    \hline
&  &  &  &  &  &  \\
Mult&($\pi^{-}$,n)40&($\pi^{-}$,p)40& 50 GeV/$c$)&Mult& 205 GeV/$c$&360 GeV/$c$ \\
&  &  &  &  &  &  \\
\hline
 n & Pn $\pm$dPn&  Pn$\pm$dPn &  Pn$\pm$dPn  & n & Pn $\pm$ dPn & Pn $\pm$ dPn  \\
\hline
 0 &      0.    &.050$\pm$.030&.007$\pm$.001 & 0 &.012$\pm$.005  & .020$\pm$.010 \\                                                                           
 1 &.100$\pm$.03&       0.    &      0.      & 2 &.069$\pm$.005  & .059$\pm$.009 \\
 2 &      0.    &.145$\pm$.025&.133$\pm$.013 & 4 &.146$\pm$.007  & .126$\pm$.009 \\
 3 &.271$\pm$.03&       0.    &      0.      & 6 &.162$\pm$.008  & .140$\pm$.009 \\
 4 &      0.    &.296$\pm$.025&.291$\pm$.015 & 8 &.171$\pm$.008  & .156$\pm$.009 \\
 5 &.259$\pm$.03&       0.    &      0.      & 10&.138$\pm$.007  & .136$\pm$.008 \\
 6 &      0.    &.263$\pm$.025&.270$\pm$.013 & 12&.091$\pm$.003  & .101$\pm$.005 \\            
 7 &.199$\pm$.04&       0.    &      0.      & 14&.046$\pm$.004  & .065$\pm$.004 \\             
 8 &      0.    &.158$\pm$.025&.181$\pm$.009 & 16&.026$\pm$.007  & .035$\pm$.002 \\
 9 &.107$\pm$.04&       0.    &      0.      & 18&.013$\pm$.002  & .020$\pm$.001 \\
 10&      0.    &.074$\pm$.020&.078$\pm$.006 & 20&.004$\pm$.002  & .010$\pm$.001 \\
 11&.05 $\pm$.03&       0.    &      0.      & 22&.0011$\pm$.0004& .004$\pm$.001 \\
 12&      0.    &.028$\pm$.009&.033$\pm$.007 & 24&.0012$\pm$.0004& .002$\pm$.001 \\                      
 13&.013$\pm$.01&       0.    &      0.      & 26&.0012$\pm$.0004&.001$\pm$.0005 \\      
 14&      0.    &.008$\pm$.004&.006$\pm$.002 & 28&        0.     &.0002$\pm$.0001 \\
 15&.003$\pm$.001&      0.    &      0.      & 30&        0.     &.0001$\pm$.0001 \\
 16&      0.    &.002$\pm$.001&.003$\pm$.0015&   &               &                \\
 17&.001$\pm$.001&      0.    &      0.      &   &               &                \\
 18&      0.    &.001$\pm$.0005&.003$\pm$.003&   &               &                \\
 19&.001$\pm$.001&        0.   &     0.      &   &               &                \\
 20&      0.    &.001$\pm$.0003&     0.      &   &               &                \\
 22&      0.    &.0002$\pm$.0002&    0.      &   &               &                \\
\hline
\end{tabular}
\   \par
\newpage
\begin{center} 
Table 9. Experimental results the multiplicity  distributions of charged particles \\  
  in (p,p) collisions at 62 GeV and at (50,200,205,400,800) GeV/$c$ [14,23-24] \\
\end{center}
\begin{tabular}{|l|c|c|c|c|c|c|}    \hline
&  &  &  &  &  &  \\
Mult& (p,p) 50   &    62 GeV   & 200 GeV/$c$)& 205 GeV/$c$ & 400 GeV/$c$  &800 GeV/$c$ \\
&  &  &  &  &  &  \\
\hline
 n & Pn$\pm$dPn  & Pn $\pm$ dPn& Pn $\pm$ dPn& Pn $\pm$ dPn&  Pn $\pm$ dPn& Pn$\pm$dPn \\
\hline                               
 2 &.158$\pm$.014&.047$\pm$.015&.075$\pm$.015&.107$\pm$.015&.082$\pm$.020&.045$\pm$.015 \\
 4 &.249$\pm$.010&.093$\pm$.015&.175$\pm$.010&.170$\pm$.007&.140$\pm$.010&.120$\pm$.020 \\
 6 &.212$\pm$.009&.103$\pm$.015&.200$\pm$.010&.212$\pm$.008&.156$\pm$.010&.150$\pm$.020 \\            
 8 &.133$\pm$.006&.113$\pm$.015&.200$\pm$.010&.177$\pm$.007&.174$\pm$.010&.160$\pm$.020 \\ 
 10&.054$\pm$.003&.115$\pm$.010&.145$\pm$.010&.135$\pm$.006&.143$\pm$.010&.150$\pm$.020 \\
 12&.013$\pm$.002&.112$\pm$.010&.100$\pm$.010&.105$\pm$.005&.116$\pm$.010&.120$\pm$.020 \\                      
 14&.005$\pm$.001&.108$\pm$.010&.005$\pm$.005&.052$\pm$.003&.085$\pm$.010&.100$\pm$.020 \\
 16&       0.    &.085$\pm$.010&.025$\pm$.005&.027$\pm$.002&.040$\pm$.010&.080$\pm$.015 \\
 18&       0.    &.065$\pm$.005&.013$\pm$.004&.010$\pm$.001&.029$\pm$.010&.040$\pm$.015 \\                
 20&       0.    &.053$\pm$.005&.005$\pm$.002&.005$\pm$.001&.017$\pm$.009&.025$\pm$.008 \\
 22&       0.    &.035$\pm$.006&        0.   &.002$\pm$.001&.010$\pm$.005&.015$\pm$.007 \\
 24&       0.    &.027$\pm$.003&             &             &.004$\pm$.002&.016$\pm$.008 \\
 26&       0.    &.020$\pm$.003&             &             &.0009$\pm$.0005&.004$\pm$.0015 \\
 28&       0.    &.010$\pm$.003&             &             &.0005$\pm$.0003&.002$\pm$.0006 \\
 30&       0.    &.006$\pm$.002&             &             &.0009$\pm$.0005&.0008$\pm$.0003 \\ 
 32&       0.    &.005$\pm$.002&             &             &               &.0006$\pm$.0003 \\
 34&       0.    &.003$\pm$.002&             &             &               &                \\
 36&       0.    &.002$\pm$.001&             &             &               &                \\
\hline
\end{tabular}
\   \par
\newpage
\begin{center} 
Table 10.Experimental results the multiplicity  distributions of charged particles \\  
 in (p,antip) collisions at (14.75[8,25], 22.4[16]) GeV/$c$ and at (200, 900) GeV[13] \\
\end{center}
\begin{tabular}{|l|c|c|c|c|c|}    \hline
&  &  &  &  &  \\
Mult& 14.75 GeV/$c$ &  22.4 GeV/$c$ &  200 GeV  & Mult &   900 GeV  \\
&  &  &  &  &  \\
\hline
 n  & Pn $\pm$ dPn   &  Pn $\pm$ dPn   &  Pn $\pm$ dPn   &  n  &  Pn $\pm$ dPn    \\
\hline
 0 & 0.050$\pm$0.030 & 0.017$\pm$0.010 &       0.        &  0  &       0.         \\     
 2 & 0.250$\pm$0.080 & 0.225$\pm$0.055 & 0.011$\pm$0.008 &  2  & 0.010$\pm$0.0015 \\
 4 & 0.080$\pm$0.025 & 0.362$\pm$0.045 & 0.045$\pm$0.009 &  4  & 0.030$\pm$0.006  \\
 6 & 0.300$\pm$0.090 & 0.242$\pm$0.024 & 0.064$\pm$0.006 &  6  & 0.060$\pm$0.006  \\            
 8 & 0.200$\pm$0.060 & 0.109$\pm$0.015 & 0.080$\pm$0.006 &  8  & 0.074$\pm$0.007  \\ 
 10& 0.050$\pm$0.020 & 0.036$\pm$0.009 & 0.100$\pm$0.010 & 14  & 0.087$\pm$0.008  \\
 12& 0.008$\pm$0.002 & 0.010$\pm$0.004 & 0.098$\pm$0.009 & 16  & 0.090$\pm$0.009  \\                      
 14& 0.001$\pm$0.0005& 0.002$\pm$0.001 & 0.102$\pm$0.008 & 24  & 0.080$\pm$0.008  \\
 16&0.0002$\pm$0.0001&0.0001$\pm$0.0001& 0.098$\pm$0.006 & 34  & 0.060$\pm$0.006  \\
 18&       0.        &         0.      & 0.094$\pm$0.006 & 40  & 0.040$\pm$0.004  \\                
 20&       0.        &         0.      & 0.086$\pm$0.006 & 48  & 0.030$\pm$0.003  \\
 22&       0.        &         0.      & 0.088$\pm$0.006 & 50  & 0.010$\pm$0.002  \\
 24&       0.        &                 & 0.076$\pm$0.007 &     &       0.         \\
 26&       0.        &                 & 0.072$\pm$0.007 &     &                  \\
 28&       0.        &                 & 0.045$\pm$0.008 &     &                  \\
 30&       0.        &                 & 0.050$\pm$0.008 &     &                  \\ 
 32&       0.        &                 & 0.028$\pm$0.009 &     &                  \\
 34&       0.        &                 & 0.010$\pm$0.005 &     &                  \\
 36&       0.        &                 & 0.005$\pm$0.002 &     &       0.         \\
\hline
\end{tabular}
\newpage
\begin{center} 
Table 11.Experimental results the multiplicity distributions of charged \\  
particles in ((p,p),(p,n),(p,d))[22],(p,Ar) and (p,Xe)$\rightarrow$($\pi^{-}$,X)[24]\\
\end{center}
\begin{tabular}{|l|c|c|c|c|c|}    \hline
Mult& (p,p) 300   & (p,n) 300    &  (p,d) 300    & (p,Ar) 200   &(p,Xe) 200 GeV/$c$ \\
&  &  &  &  &  \\
 n & Pn $\pm$ dPn &  Pn $\pm$dPn &  Pn $\pm$dPn  & Pn $\pm$ dPn & Pn $\pm$ dPn \\
 0 &      0.      &        0.    &       0.      &        0.    &.025$\pm$.015 \\                                                                           
 1 &      0.      &.061$\pm$.015 &.040$\pm$0.015 &        0.    &       0.     \\
 2 &.063$\pm$0.015&        0.    &.061$\pm$0.015 &.027$\pm$.008 &.060$\pm$.020 \\
 3 &      0.      &.121$\pm$.012 &.110$\pm$0.015 &        0.    &.080$\pm$.020 \\
 4 &.130$\pm$0.010&        0.    &.130$\pm$0.015 &.035$\pm$.006 &.075$\pm$.020 \\
 5 &      0.      &.138$\pm$.011 &       0.      &        0.    &.115$\pm$.020 \\
 6 &.139$\pm$0.011&        0.    &.120$\pm$0.010 &        0.    &.117$\pm$.020 \\            
 7 &      0.      &.147$\pm$.015 &       0.      &.041$\pm$.006 &.110$\pm$.020 \\             
 8 &.161$\pm$0.015&        0.    &       0.      &        0.    &.075$\pm$.020 \\
 9 &      0.      &.130$\pm$.013 &.094$\pm$0.007 &        0.    &.060$\pm$.020 \\
 10&.135$\pm$0.014&        0.    &       0.      &.057$\pm$.006 &.052$\pm$.020 \\
 11&      0.      &.107$\pm$.008 &       0.      &        0.    &.045$\pm$.010 \\
 12&.101$\pm$0.013&        0.    &.073$\pm$0.007 &        0.    &.035$\pm$.008 \\                      
 13&      0.      &.067$\pm$.007 &       0.      &.045$\pm$.006 &.025$\pm$.007 \\      
 14&.062$\pm$0.010&        0.    &       0.      &        0.    &.020$\pm$.006 \\
 15&      0.      &.044$\pm$.006 &.058$\pm$0.007 &        0.    &.015$\pm$.006 \\
 16&.036$\pm$0.008&        0.    &       0.      &.038$\pm$.006 &.013$\pm$.006 \\
 17&      0.      &.018$\pm$.005 &       0.      &        0.    &.012$\pm$.006 \\
 18&.010$\pm$0.005&        0.    &.048$\pm$0.007 &        0.    &.011$\pm$.006 \\               
 19&      0.      &.006$\pm$.002 &        0.     &.029$\pm$.006 &.010$\pm$.006 \\
 20&.009$\pm$0.004&        0.    &.029$\pm$0.005 &        0.    &.010$\pm$.006 \\                 
 21&      0.      &.0063$\pm$.003&.015$\pm$0.006 &        0.    &.010$\pm$.006 \\
 22&.003$\pm$.0015&        0.    &.006$\pm$0.003 &.028$\pm$.006 &.012$\pm$.007 \\
 23&      0.      &.002$\pm$.001 &        0.     &        0.    &        0.    \\  
 24&.002$\pm$0.001&        0.    &        0.     &        0.    &        0.    \\
 25&      0.      &        0.    &        0.     &.011$\pm$.006 &        0.    \\
 27&      0.      &        0.    &        0.     &        0.    &        0.    \\
 28&      0.      &        0.    &        0.     &.016$\pm$.006 &        0.    \\
 29&      0.      &        0.    &        0.     &        0.    &        0.    \\
 31&      0.      &        0.    &        0.     &.009$\pm$.004 &        0.    \\
 34&      0.      &        0.    &        0.     &.008$\pm$.004 &        0.    \\
 37&      0.      &        0.    &        0.     &.007$\pm$.004 &        0.    \\
 40&      0.      &        0.    &        0.     &.006$\pm$.003 &        0.    \\
\hline
\end{tabular}
\newpage
\begin{center} 
Table 12. Experimental results the multiplicity distributions of charged \\  
particles in (p,Xe) and (p,Ar)$\rightarrow$($\pi^{-}$,X) collisions [24] \\
\end{center}
\begin{tabular}{|l|c|c|c|c|c|}    \hline
&  &  &  &  &  \\
Mult& (p,Xe)ch 200 & Mult & (p.Xe)ch 200 & Mult & (p,Ar)$\rightarrow$($\pi^{-}$,X) 200 GeV/$c$ \\
\hline
&  &  &  &  &  \\
 n & Pn $\pm$ dPn  & n   &  Pn $\pm$dPn   & n    &  Pn $\pm$dPn    \\
\hline
 0 &      0.       &  31 &0.014$\pm$0.006 &   0  & 0.025$\pm$0.015 \\                                                                           
 1 &      0.       &  34 &0.017$\pm$0.006 &   1  &         0.      \\
 2 &0.019$\pm$0.008&  37 &0.011$\pm$0.005 &   2  & 0.080$\pm$0.020 \\
 3 &      0.       &  40 &0.011$\pm$0.005 &   3  & 0.100$\pm$0.020 \\
 4 &0.027$\pm$0.007&  43 &0.011$\pm$0.005 &   4  & 0.130$\pm$0.020 \\
 5 &      0.       &  46 &0.008$\pm$0.004 &   5  & 0.150$\pm$0.020 \\
 6 &      0.       &  49 &0.007$\pm$0.004 &   6  & 0.120$\pm$0.020 \\            
 7 &0.035$\pm$0.007&  52 &0.005$\pm$0.003 &   7  & 0.100$\pm$0.020 \\             
 8 &      0.       &  55 &0.004$\pm$0.003 &   8  & 0.070$\pm$0.020 \\
 9 &      0.       &  58 &0.003$\pm$0.002 &   9  & 0.050$\pm$0.020 \\
 10&0.033$\pm$0.006&  60 &0.003$\pm$0.002 &   10 & 0.035$\pm$0.010 \\
 11&      0.       &     &  0.            &   11 & 0.030$\pm$0.010 \\
 12&      0.       &     &  0.            &   12 & 0.025$\pm$0.006 \\                      
 13&0.035$\pm$0.006&     &  0.            &   13 & 0.013$\pm$0.006 \\      
 14&      0.       &     &  0.            &   14 & 0.011$\pm$0.006 \\
 15&      0.       &     &  0.            &   15 & 0.010$\pm$0.006 \\
 16&0.033$\pm$0.006&     &  0.            &   16 & 0.011$\pm$0.006 \\
 17&      0.       &     &  0.            &      &                 \\
 18&      0.       &     &  0.            &      &                 \\               
 19&0.027$\pm$0.006&     &  0.            &      &                 \\
 20&      0.       &     &  0.            &      &                 \\                 
 21&      0.       &     &  0.            &      &                 \\
 22&0.023$\pm$0.006&     &  0.            &      &                  \\
 23&      0.       &     &  0.            &      &                  \\  
 24&      0.       &     &  0.            &      &                  \\
 25&0.018$\pm$0.006&     &  0.            &      &                  \\
 26&      0.       &     &  0.            &      &                  \\
 27&      0.       &     &  0.            &      &                  \\
 28&0.016$\pm$0.006&     &  0.            &      &                  \\
 29&      0.       &     &  0.            &      &                  \\
 30&      0.       &     &  0.            &      &                  \\
\hline
\end{tabular} 
\newpage 
\begin{figure}
\begin{center}
\epsfig{file=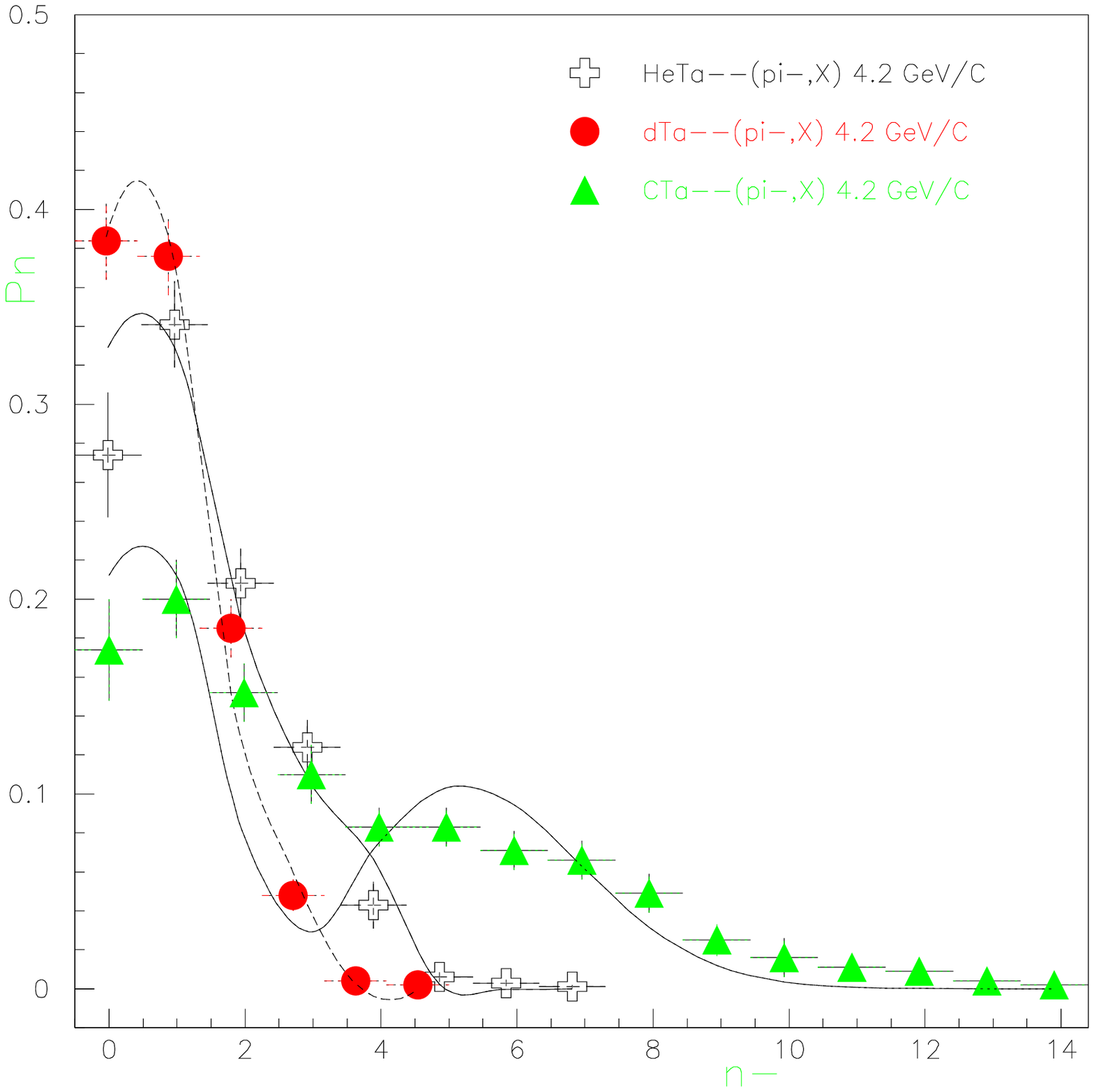,bbllx=0pt,bblly=0pt,bburx=594pt,bbury=842pt,
width=18cm,angle=0}
\end{center}
\vspace{-5.5cm}
\begin{minipage}{15.0cm}
\caption
{The multiplicity  distributions of  $\pi^{-}$ 
mesons in ((He,d,C),Ta) collisions at 4.2 GeV/$c$/nucleon.}
\end{minipage}
\end{figure}
\begin{figure}
\begin{center}
\epsfig{file=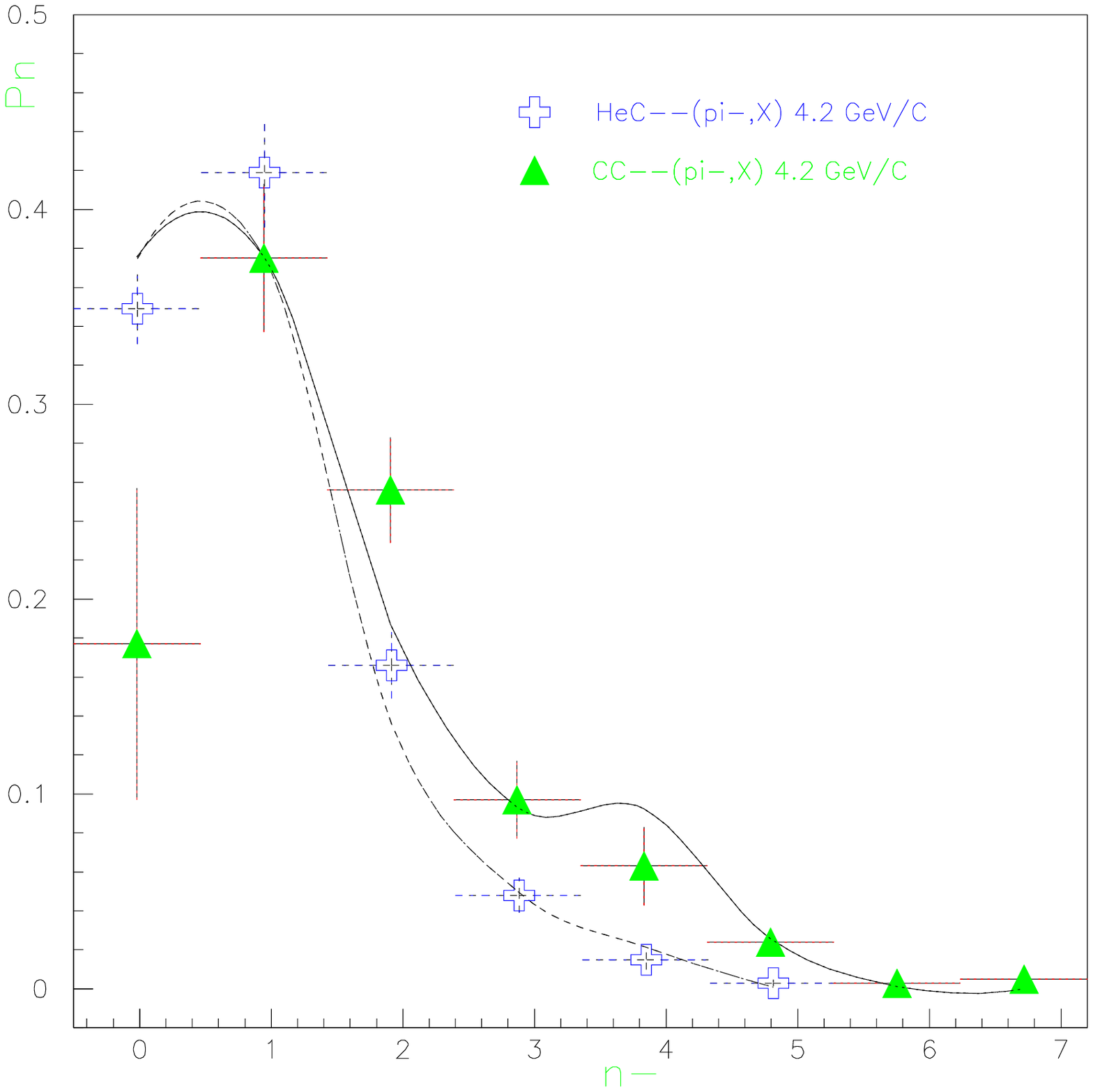,bbllx=0pt,bblly=0pt,bburx=594pt,bbury=842pt,
width=18cm,angle=0}
\end{center}
\vspace{-5.5cm}
\begin{minipage}{15.0cm}
\caption
{The multiplicity  distributions of  $\pi^{-}$ 
mesons in ((He,C),C) collisions at 4.2 GeV/$c$/nucleon.}
\end{minipage}
\end{figure}
\begin{figure}
\begin{center}
\epsfig{file=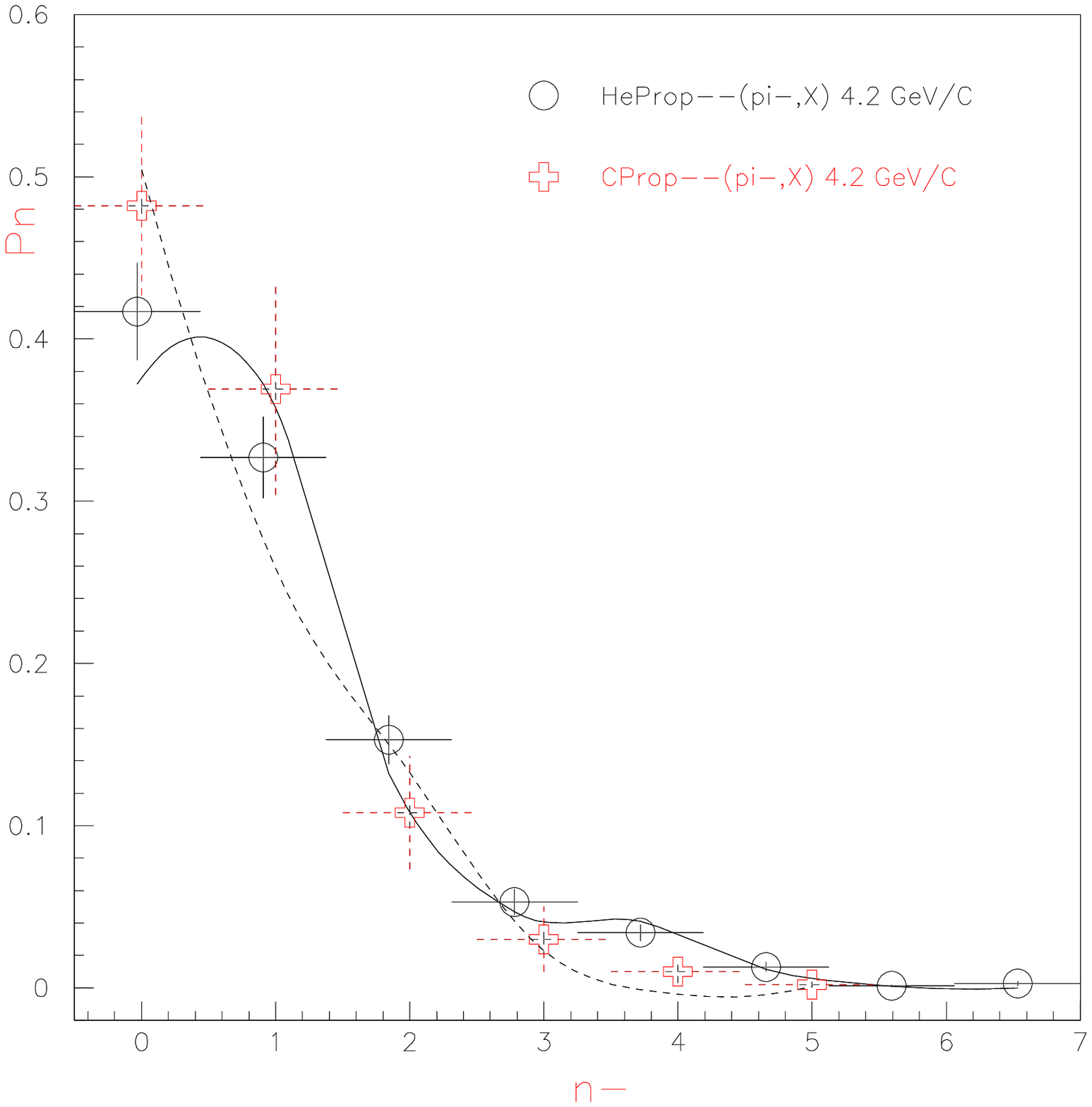,bbllx=0pt,bblly=0pt,bburx=594pt,bbury=842pt,
width=18cm,angle=0}
\end{center}
\vspace{-5.5cm}
\begin{minipage}{15.0cm}
\caption
{The multiplicity distributions of  $\pi^{-}$ mesons in ((He,C),Prop)
 collisions at 4.2 GeV/$c$/nucleon.}
\end{minipage}
\end{figure}
\begin{figure}
\begin{center}
\epsfig{file=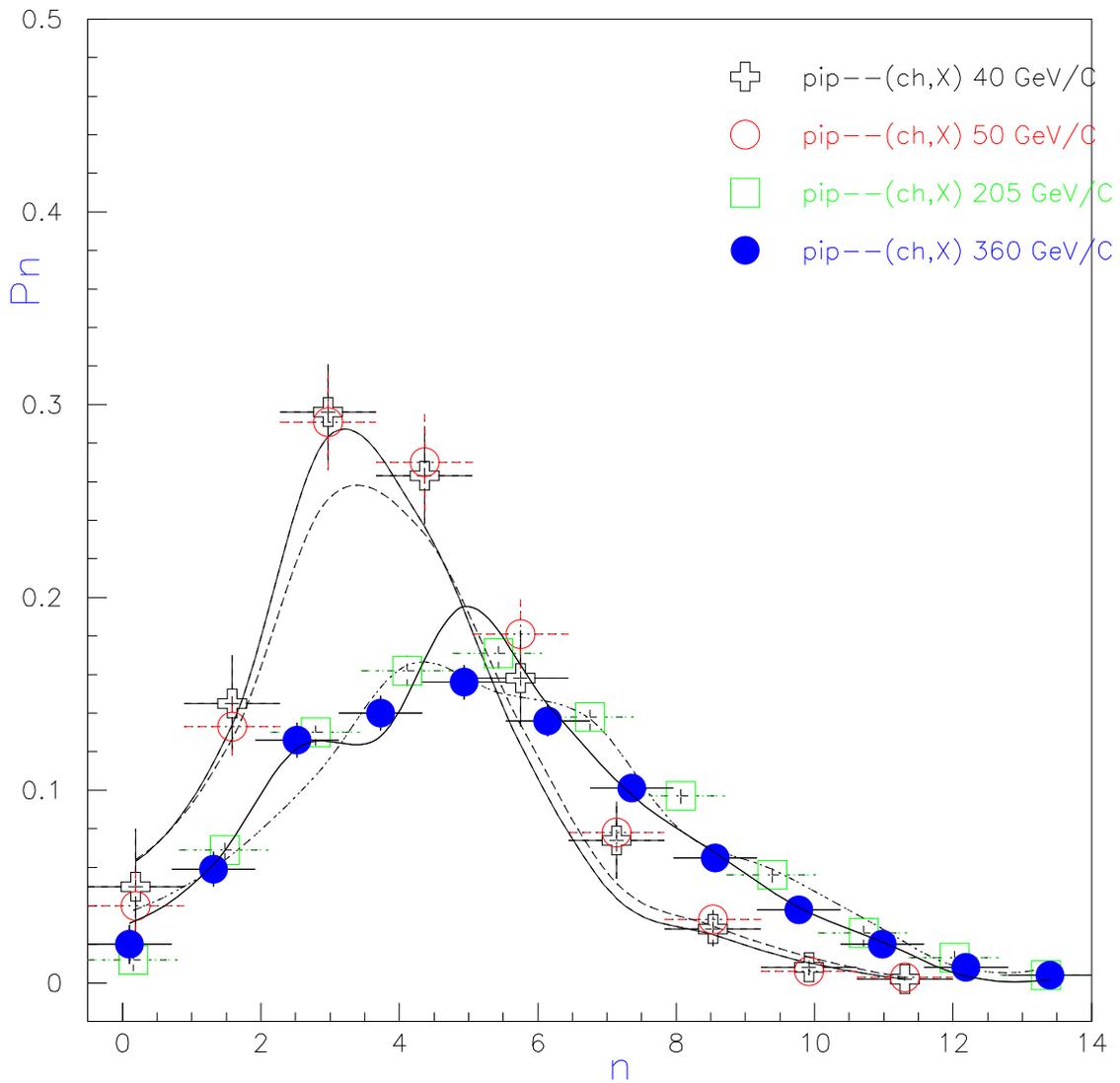,bbllx=0pt,bblly=0pt,bburx=594pt,bbury=842pt,
width=18cm,angle=0}
\end{center}
\vspace{-5.5cm}
\begin{minipage}{15.0cm}
\caption
{The multiplicity distributions of charged particles in 
($\pi^{-}$,p)$\rightarrow$(ch,X) at (40, 50, 205 and 360) GeV/$c$.}
\end{minipage}
\end{figure}
\begin{figure}
\begin{center}
\epsfig{file=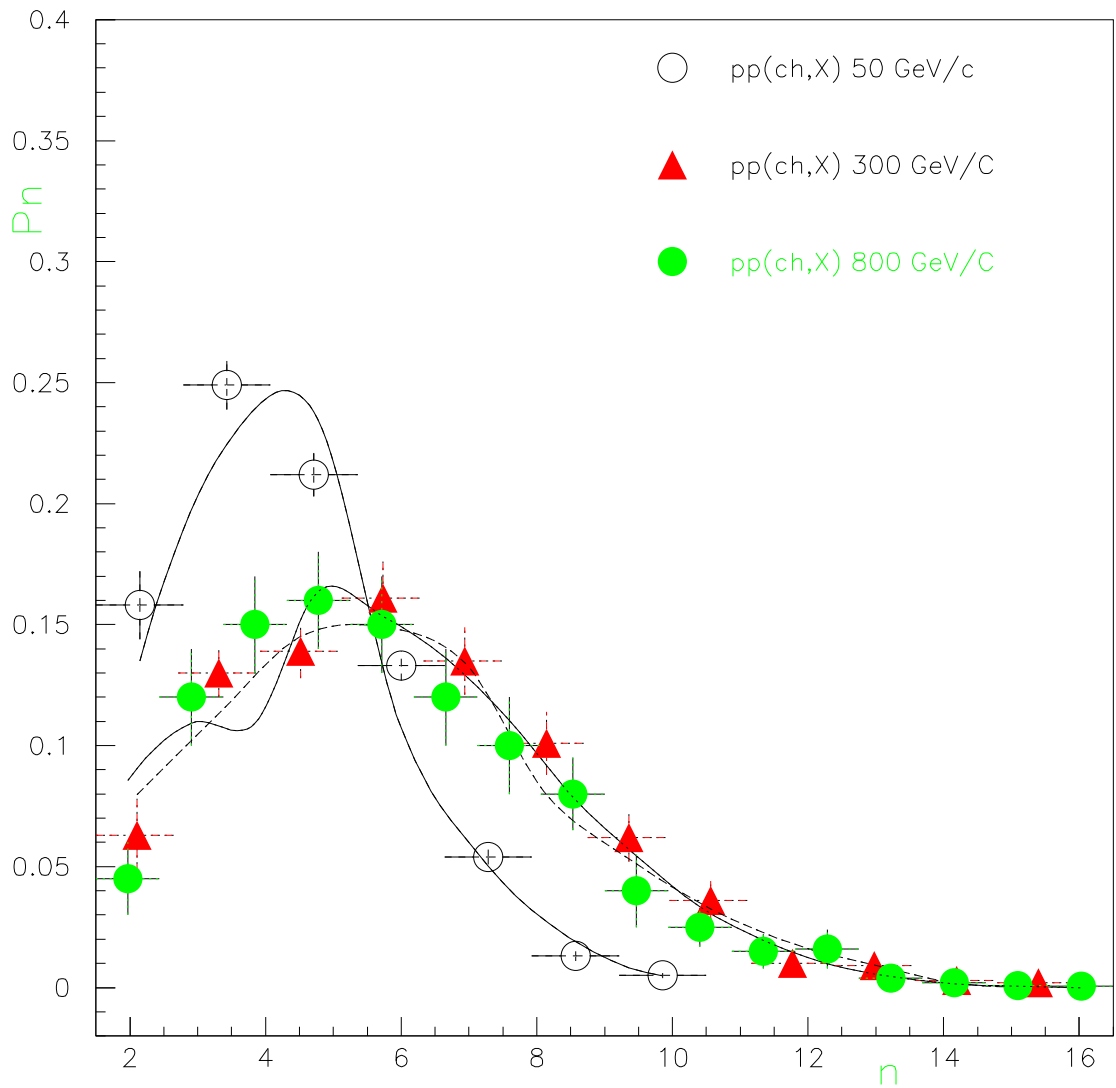,bbllx=0pt,bblly=0pt,bburx=594pt,bbury=842pt,
width=18cm,angle=0}
\end{center}
\vspace{-5.5cm}
\begin{minipage}{15.0cm}
\caption
{The multiplicity  distributions of charged particles in 
(p,p)$\rightarrow$(ch,X) at (50,300 and 800) GeV/$c$. }
\end{minipage}
\end{figure}
\begin{figure}
\begin{center}
\epsfig{file=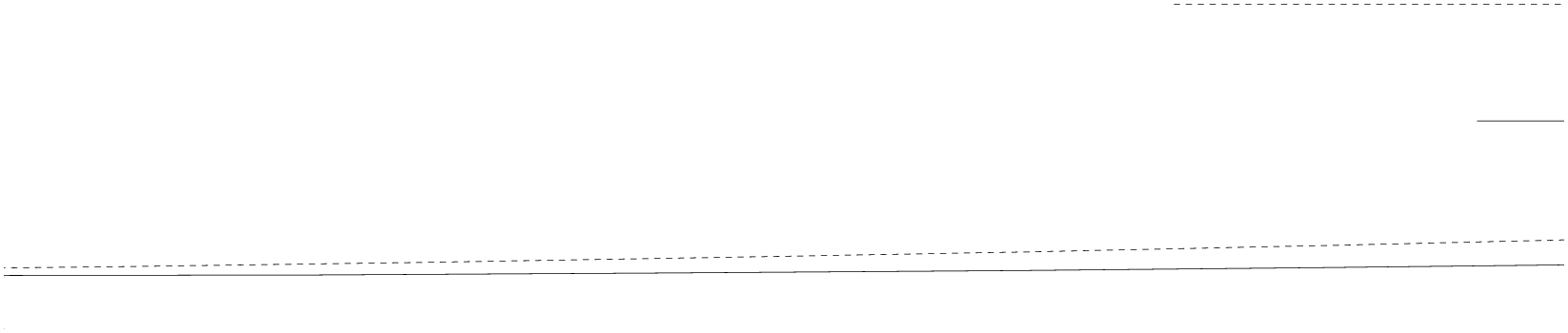,bbllx=0pt,bblly=0pt,bburx=594pt,bbury=842pt,
width=18cm,angle=0}
\end{center}
\vspace{-5.5cm}
\begin{minipage}{15.0cm}
\caption
{The multiplicity distributions of charged particles in 
(p,antip)$\rightarrow$(ch,X) at (14.75 and 22.4) GeV/$c$.}
\end{minipage}
\end{figure}
\begin{figure}
\begin{center}
\epsfig{file=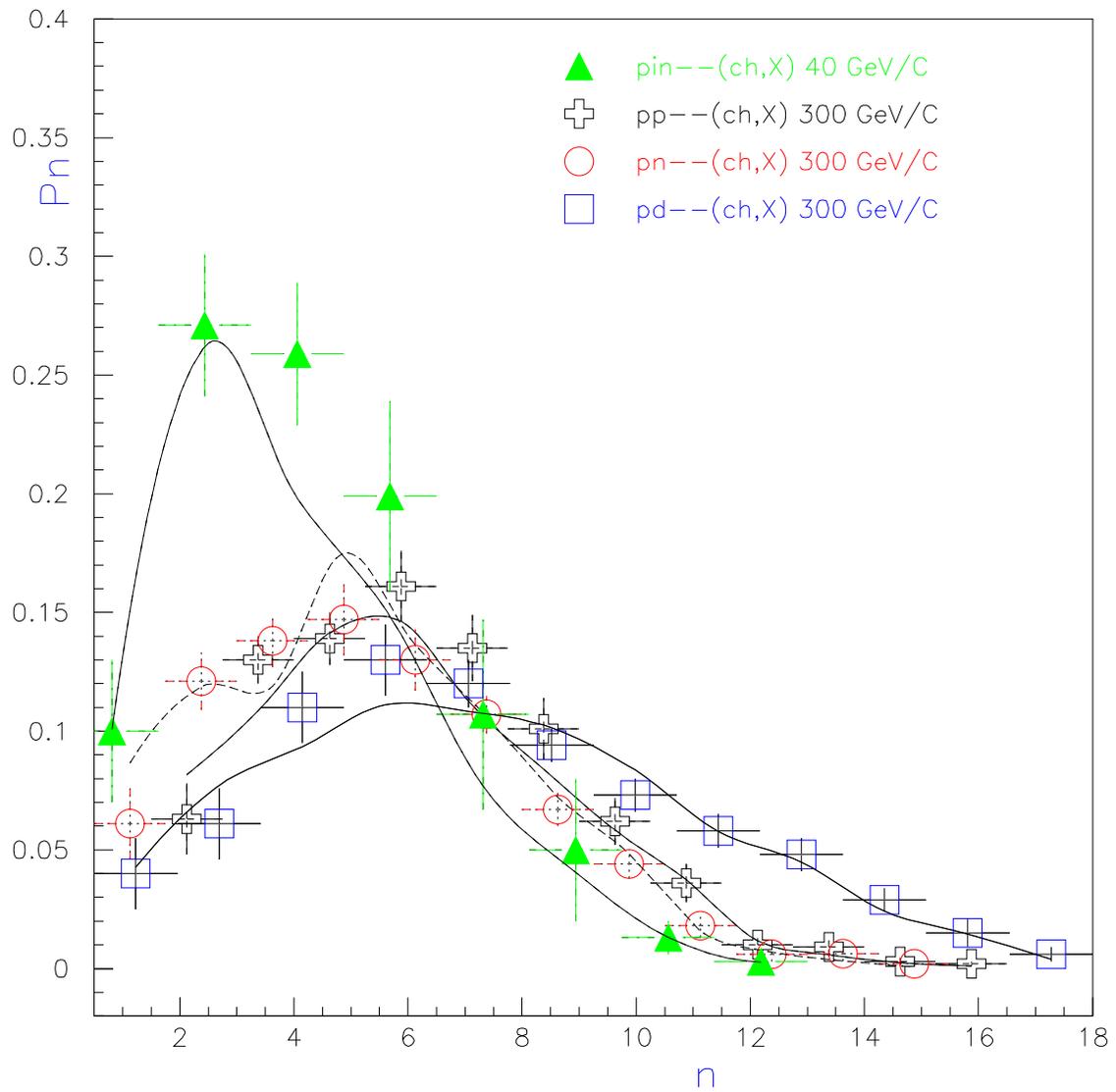,bbllx=0pt,bblly=0pt,bburx=594pt,bbury=842pt,
width=18cm,angle=0}
\end{center}
\vspace{-5.5cm}
\begin{minipage}{15.0cm}
\caption
{The multiplicity distributions of charged particles ($\pi^{-}$,n)$\rightarrow$(ch,X) 
 at the momentum of 40 GeV/$c$/nucleon and
 (p,(p,n,d))$\rightarrow$ (ch,X) at the momentum of 300 GeV/$c$/nucleon.}
\end{minipage}
\end{figure}
\begin{figure}
\begin{center}
\epsfig{file=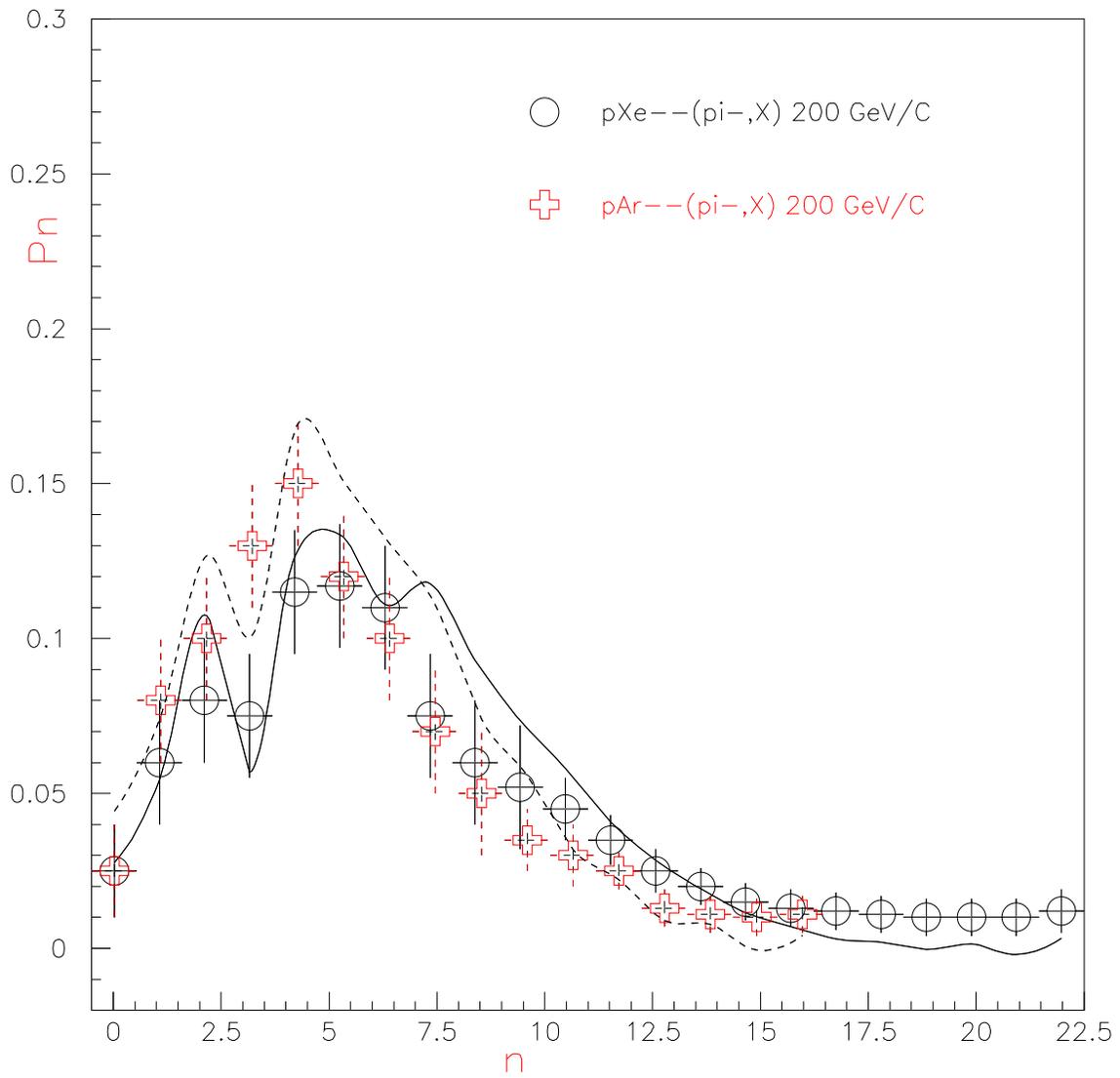,bbllx=0pt,bblly=0pt,bburx=594pt,bbury=842pt,
width=18cm,angle=0}
\end{center}
\vspace{-5.5cm}
\begin{minipage}{15.0cm}
\caption
{The multiplicity  distributions of $\pi^{-}$ mesons 
 in (p,(Ar,Xe))$\rightarrow$($\pi^{-}$,X) 
 collisions at 200 GeV/$c$/nucleon. }
\end{minipage}
\end{figure}
\begin{figure}
\begin{center}
\epsfig{file=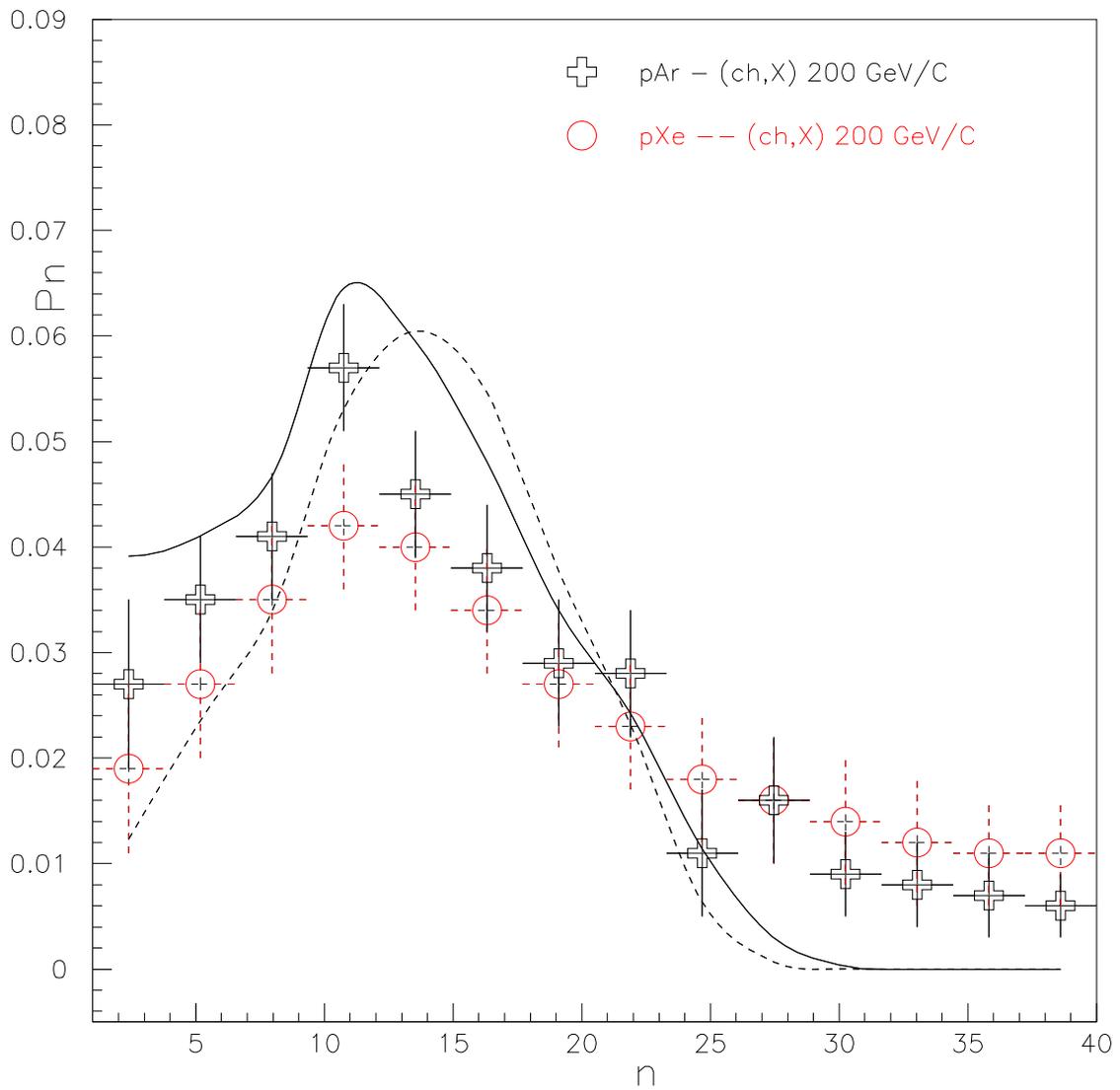,bbllx=0pt,bblly=0pt,bburx=594pt,bbury=842pt,
width=18cm,angle=0}
\end{center}
\vspace{-5.5cm}
\begin{minipage}{15.0cm}
\caption
{The multiplicity  distributions of charged particles  
 in (p,(Ar,Xe))$\rightarrow$(ch,X) 
 collisions at 200 GeV/$c$/nucleon. }
\end{minipage}
\end{figure}
\end{document}